\documentclass{article}
\usepackage{arxiv}
\usepackage[utf8]{inputenc} 
\usepackage[T1]{fontenc}    
\usepackage{hyperref}       
\usepackage{url}            
\usepackage{booktabs}       
\usepackage{amsfonts}       
\usepackage{nicefrac}       
\usepackage{microtype}      
\usepackage{lipsum}		
\usepackage{graphicx}
\usepackage[square,numbers]{natbib}
\usepackage{doi}
\usepackage{amssymb}
\usepackage{bm}
\usepackage{amstext}
\usepackage{amsmath}
\usepackage{xcolor}

\title{Impact of reconstruction schemes on interpreting lattice Boltzmann results\\A study using the Taylor-Green vortex problem}

\author{ \href{https://orcid.org/0000-0001-9913-5180}{\includegraphics[scale=0.06]{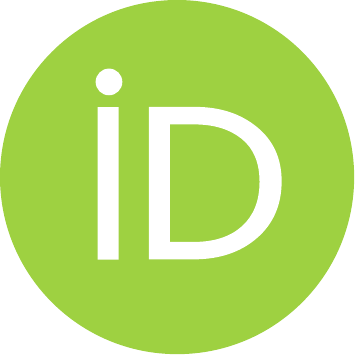}\hspace{1mm}Jianping Meng},  \href{https://orcid.org/0000-0001-9310-1465}{\includegraphics[scale=0.06]{orcid.pdf}\hspace{1mm}Xiao-Jun Gu},   \href{https://orcid.org/0000-0002-6085-5049}{\includegraphics[scale=0.06]{orcid.pdf}\hspace{1mm}David R. Emerson} \\
Scientific Computing Department, STFC Daresbury Laboratory, Warrington, United Kingdom, WA4 4AD\\
}

\renewcommand{\headeright}{}
\renewcommand{\undertitle}{}
\renewcommand{\shorttitle}{Impact of reconstruction schemes on interpreting lattice Boltzmann results}

\hypersetup{
pdftitle={Impact of reconstruction schemes on interpreting lattice Boltzmann results},
pdfauthor={Jianping Meng, Xiao-Jun Gu, David R. Emerson},
pdfkeywords={Lattice Boltzmann method, Numerical accuracy, Taylor-Green vortex},
}

\begin{document}
\maketitle

\begin{abstract}
	In this note, we show how reconstruction schemes can have a significant impact on interpreting lattice Boltzmann simulation data. To reconstruct turbulence quantities, e.g., the kinetic energy dissipation rate and enstrophy, schemes higher than second-order for spatial derivatives can greatly improve the prediction of these quantities in the Taylor-Green vortex problem. In contrast, a second-order reconstruction of the time series data indicates very good accuracy for the kinetic energy dissipation rate. The present findings can be considered as further numerical evidence of the capability of the lattice Boltzmann method to simulate turbulent flows, which is consistent with its proven feature of low numerical diffusion.
\end{abstract}


\section{Introduction}\label{sec:introduction}

Numerical modeling is an important approach for studying complex fluid behavior occurring in both nature and industrial processes. For this purpose, a key step is to design accurate and simple numerical schemes. As an example, low numerical dissipation and dispersion are often desired features for many applications, e.g., direct numerical simulation of turbulence and aeroacoustics~\cite{Ekaterinaris2005}.

The lattice Boltzmann method (LBM) is an emerging tool suitable for modeling many types of fluid flow~\cite{chen_lattice_1998}. The method originates from lattice gas automata (LGA)~\cite{Wolfram1986,frisch_lattice-gas_1986}, and was subsequently identified as a special discrete velocity method for the Boltzmann transport equation (BTE)~\cite{he_priori_1997,he_theory_1997,shan_discretization_1998,shan_kinetic_2006}. Only a minimal set of discrete velocities are kept to retrieve part of the BTE solution domain, typically incompressible or weakly compressible flows at continuum and near-continuum regimes. In addition, it also inherits the simplicity of the LGA, i.e., its particle dynamics algorithm that only requires information from its near neighbors with fixed directions (i.e., no neighborhood search is required). Thus, the method provides an ideal platform for simulating complex fluid phenomena~\cite{aidun_lattice-boltzmann_2010} and is ideal for high-performance computing without worrying about complex numerical and computational details.

The Chapman-Enskog expansion proves that the incompressible Navier-Stokes (NS) equations are a first-order asymptotic solution, in terms of the Knudsen number (i.e., approaching the continuum limit), of the lattice Boltzmann equation. In particular, the numerical diffusion can be absorbed into the physical viscosity, which leads to very low numerical diffusion error. In~\cite{marie_comparison_2009}, it was revealed that LBM demonstrates remarkable lower numerical diffusion and dispersion error, even compared with high-order numerical discretization of the NS equations. This finding indicates that the numerical accuracy of LBM at the macroscopic scale might not be strictly correlated or limited to its formal second-order accuracy measured relative to the BTE.

Recently, its accuracy has been numerically assessed against the classical Taylor-Green vortex (TGV) and other benchmark problems associated with turbulent flow regimes~\cite{Mimeau2021,Haussmann2019,Nathen2018}. In general, satisfactory accuracy is identified, although there are differences among various collision terms, such as the single-relaxation model, the multiple-relaxation model, and the entropic collision model etc. For the TGV problem, the kinetic energy can be predicted with fairly good accuracy using a mesh of \(128^3\). However, Mimeau \textit{et al}. observed that a relatively high mesh resolution is required to predict the enstrophy~\cite{Mimeau2021}, where spatial derivatives of velocity are needed.  Interestingly, better accuracy, although with oscillations, is found for the kinetic energy dissipation rate reconstructed from the kinetic energy time series.  However, there should be only a constant factor of difference between the enstrophy and the kinetic energy dissipation rate  if they are evaluated using spatial derivatives for this particular problem.

In LBM simulations, the evolutionary variables are distribution functions and not macroscopic quantities, like density and velocity. Thus, the kinetic energy dissipation rate and enstrophy are purely reconstructed from the calculated velocity data. More specifically, evaluating macroscopic velocity-gradient terms are not involved in LBM simulations, and the reconstruction process is totally independent from the simulation itself. Thus, in principle, there is no restriction on the choice of reconstruction scheme. In previous work, however, the reconstruction process is often conducted with second-order schemes by default e.g., in~\cite{Mimeau2021}.

In this work, we investigate the impact of the reconstruction scheme on interpreting lattice Boltzmann simulation results, which may explain the reason for different accuracy observed in previous predictions of the kinetic energy dissipation rate and enstrophy.  For this purpose, we employ the TGV problem and use various reconstruction schemes based on the same set of simulation data. In the following, we briefly introduce the lattice Boltzmann method and the Taylor-Green vortex problem, and discuss the impact of reconstruction schemes.

\section{Lattice Boltzmann method}\label{sec:lattice-boltzmann-method}

To derive the LBM, we can start from the Boltzmann-BGK type kinetic equation~\cite{he_priori_1997,he_theory_1997,shan_discretization_1998,shan_kinetic_2006},

\begin{equation}
    \frac{\partial f}{\partial t}+\bm{C}\cdot\nabla f=-\frac{1}{\tau}\left(f-f^{eq}\right),\label{continouspde}
\end{equation}
which describes fluid flows using the distribution function \(f(\bm{r},\bm{C},t)\) at position
\(\bm{r}=(x,y,z)\), molecular velocity \(\bm{C}=(C_x,C_y,C_z)\), and time,  \(t\). The molecular collision is modeled by a relaxation term towards the equilibrium distribution function

\begin{equation}
    f^{eq}=\rho \left(\frac{1}{2\pi R T}\right)^{D/2} \exp\left[-\frac{(\bm{C}-\bm{U})^2}{2RT}\right] \label{bgkcoll}
\end{equation}
which is determined by the fluid density, $\rho$, the fluid velocity, \(\bm{U}=(u,v,w)\), and the temperature, $T$. In the continuum and near-continuum limit, Eqs.~(\ref{continouspde}) and ~(\ref{bgkcoll}) can recover the NS equations via the first-order Chapman-Enksog expansion, i.e., the NS equations are a type of asymptotic solution of the kinetic equation in terms of the Knudsen number. In particular, $\tau$, is related to the fluid viscosity, \(\mu\), and the pressure, $p$, via the Chapman-Enskog expansion, i.e., \(\mu=p\tau\). Naturally, the Reynolds number
can be defined as \(Re=\rho_{0}U_{0}L/\mu_0=U_{0}L/(\tau_0 RT_{0})\), where the subscript $0$ denotes the reference value and $L$ the characteristic length of the system. The ideal gas law, \(p_0=\rho_0 R T_0\), is used for deriving the form of the Reynolds number. The Knudsen number
can also be written as \(\mu_{0}\sqrt{RT_{0}}/(p_{0}L)\). If \(\sqrt{RT_0}\) is conveniently considered as the sound speed, although there is a constant factor difference, we have \(Kn=Re/Ma\) where \(Ma=U_0/\sqrt{RT_0}\). With Eq.~(\ref{bgkcoll}), however, the thermal conductivity is not independent of the viscosity, which can be corrected by using other kinetic equations if necessary.

Since there are \(6+1\) degrees of freedom, i.e., the physical space, \(\bm{r}\), the molecular velocity space, \(C\), and the time,  Eq.~(\ref{continouspde}) can be very expensive for numerical solutions. However, for a broad range of flow problems, it is possible to greatly reduce the complexity by truncating the Maxwellian and discretizing the molecular velocity space by using Gauss-Hermite quadrature. After the discretization in the molecular velocity space, Eq.~(\ref{continouspde}) becomes
\begin{equation}
    \frac{\partial f_{\alpha}}{\partial t}+\bm{C}_{\alpha}\cdot\nabla f_{\alpha}=-\frac{1}{\tau}(f_{\alpha}-f_{\alpha}^{eq}),\label{pde}
\end{equation} where \(\bm{C}_\alpha\) is an abscissa of a Gauss–Hermite quadrature. For simulating incompressible
and isothermal flows, it is common to use second-order truncation of the Maxwellian function, i.e.,

\begin{equation}
    f_{\alpha}^{eq}=w_{\alpha}\rho\left[1+\frac{\bm{U}\cdot\bm{C}_{\alpha}}{RT_{0}}+\frac{1}{2}\frac{(\bm{U}\cdot\bm{C}_{\alpha})^{2}}{(RT_{0})^{2}}-\frac{\bm{U}\cdot\bm{U}}{2RT_{0}]}\right],\label{feq}
\end{equation}
where $w_{\alpha}$ is the weight of the Gauss-Hermite quadrature. Moreover, the density and velocity are now calculated by summation operations, i.e.,

\begin{equation}
    \rho=\sum_{\alpha}f_{\alpha}=\sum_{\alpha}f_{\alpha}^{eq},\label{calcrho}
\end{equation} and
\begin{equation}
    \rho\bm{U}=\sum_{\alpha}f_{\alpha}\bm{C}_{\alpha}=\sum_{\alpha}f_{\alpha}^{eq}\bm{C}_{\alpha}.\label{calcu}
\end{equation}
With the obtained density, the pressure can be calculated using \(p=\rho R T_0\) in isothermal flows since the temperature is constant, i.e., $T_0$.

To numerically solve Eq.~(\ref{pde}), a trapezoidal scheme in time \citep{he_novel_1998},
 can be employed at the right hand side and the left hand side can be analytically integrated over time. In particular,
by introducing

\begin{equation}
    \tilde{f}_{\alpha}=f_{\alpha}+\frac{dt}{2\tau}(f_{\alpha}-f_{\alpha}^{eq}),
\end{equation} the implicitness of the trapezoidal scheme can be eliminated and we obtain an explicit scheme

\begin{equation}
    \tilde{f}_{\alpha}(\bm{r}+\bm{C}_{\alpha}dt,t+dt)-\tilde{f}_{\alpha}(\bm{r},t)=-\frac{dt}{\tau+0.5dt}\left[\tilde{f}_{\alpha}(\bm{r},t)-f_{\alpha}^{eq}(\bm{r},t)\right],\label{eq:scheme}
\end{equation} which is ready for implementing the stream-collision scheme.
At the same time, the macroscopic
quantities can be calculated as

\begin{equation}
    \rho=\sum_{\alpha}\tilde{f}_{\alpha},\ \mbox{{and},\ }\rho\bm{U}=\sum_{\alpha}\bm{C}_{\alpha}\tilde{f_{\alpha}}.\label{eq:hemac}
\end{equation}
For three-dimensional flows, there are a few commonly used lattices, e.g., D3Q15, D3Q19 and D3Q27. Here, we choose the D3Q19 lattice where there are nineteen discrete velocities ($\alpha=1..19$),

\begin{equation}
    C_{\alpha,x}=\sqrt{3RT_{0}}[0, -1, 0, 0, -1, -1, -1, -1, 0, 0, 1, 0,  0, 1, 1,  1,  1,  0,  0],
\end{equation}
\begin{equation}
    C_{\alpha,y}=\sqrt{3RT_{0}}[0, 0, -1, 0, -1, 1, 0, 0, -1, -1, 0, 1, 0,  1, -1, 0, 0, 1, 1],
\end{equation}
\begin{equation}
    C_{\alpha,z}=\sqrt{3RT_{0}}[0, 0, 0, -1, 0, 0, -1, 1, -1, 1, 0, 0, 1, 0,  0, 1, -1, 1, -1],
\end{equation}
and the corresponding weights \(w_{\alpha}\) are

\begin{equation}
    [\frac{1}{3}, \frac{1}{18}, \frac{1}{18}, \frac{1}{18}, \frac{1}{36}, \frac{1}{36}, \frac{1}{36}, \frac{1}{36}, \frac{1}{36}, \frac{1}{36},
        \frac{1}{18}, \frac{1}{18}, \frac{1}{18}, \frac{1}{36}, \frac{1}{36}, \frac{1}{36}, \frac{1}{36}, \frac{1}{36}, \frac{1}{36}].
\end{equation}

As previously discussed, it is now ready to implement the stream-collision algorithm. The only trick is to tie the space and time step
together as $d\bm{r}=\bm{C}_{\alpha}dt$~\cite{chen_lattice_1998,shan_kinetic_2006}. For instance, assuming the
system length is \(\mathcal{L}\), we may set the spatial step \(dx=\mathcal{L}/N\) and then
$dt=\mathcal{L}/(N\sqrt{3RT_{0}})$, where $N$ is the number of computational cells. This
ensures the ``particles'' are jumping on a uniform grid system.

Although the LBM algorithm is very simple, there are two important features. First, the mass and momentum conservation are kept exactly at mesoscale (numerically within machine precision), cf. Eqs.~(\ref{calcrho}) and ~(\ref{calcu}), which might lead to highly accurate solutions of density and velocity. Second, Eq.~(\ref{eq:scheme}) is a second-order scheme in both time and space for Eq.~(\ref{continouspde}) or Eq.~(\ref{pde}), which recovers the NS equations without error in viscosity through the Chapman-Enskog expansion.  This feature can also be proved if the algorithm is derived from the lattice gas automata, cf.~\cite{chen_lattice_1998}.  Thus, the LBM simulation tends to produce very low numerical diffusion error at the macroscopic level.

 As has been shown, macroscopic quantities such as density, \(\rho\), and velocity, \(\bm{U}\), are not primitive variables in LBM simulations. Moreover, the non-linear advection term in the NS equations is replaced with a molecular streaming process, and the momentum diffusivity automatically emerges from the collective behavior of molecular collisions~\cite{Succi2018}. Hence, no gradient terms are evaluated during simulations. The macroscopically defined turbulence quantities, like enstrophy, can only be reconstructed through the extracted macroscopic velocity data. As shown in~\cite{marie_comparison_2009}, the second-order accuracy of Eq.~(\ref{eq:scheme}) at mesoscopic level might not be directly correlated with the accuracy at the macroscopic level. Consequently, reconstruction schemes, which may have a great impact on the accuracy, can be chosen not necessarily following the accuracy order of Eq.~(\ref{eq:scheme}) in space and time.

\section{3D Taylor-Green vortex and turbulence quantities}\label{sec:3d-taylor-green-vortex}

The 3D Taylor-Green vortex is a classical problem for evaluating the performance of a numerical method, particularly the numerical diffusion caused by discretization schemes. It belongs to a class of decaying homogeneous isotropic turbulence, where the flow is defined in a three-dimensional periodic box of length \(2\pi L\)  and initialized with

\begin{equation}
    \bm U(0)=\begin{pmatrix}
        U_0 \sin(\frac{x}{L}) \cos(\frac{y}{L})\cos(\frac{z}{ }) \\
        -U_0 \cos(\frac{x}{L}) \sin(\frac{y}{ L})\cos(\frac{z}{L}) \\
        0
    \end{pmatrix},
\end{equation}
and

\begin{equation}
    p(0)=p_\infty+\frac{\rho_0 U^2_0}{16}\left[\cos\left(\frac{2x}{ L}\right)+\cos\left(\frac{2y}{ L}\right)\right]\left[\cos\left(\frac{2z}{L}\right)+2\right],
\end{equation}
where \(L\) is a  characteristic length scale. In LBM simulations, any macroscopic initial condition must be transformed into the form of the distribution function. For this purpose, we employ the first-order Chapman-Enskog solution,

\begin{equation}
    f_\alpha(0)=f_\alpha^{eq}\left[\rho(0),\bm{U}(0)\right]-\tau f_\alpha^{eq}\left[\rho(0),\bm{U}(0)\right] \left[\bm{C}_\alpha \bm{C}_\alpha:\bm{\nabla} \bm{U\ }(0)\right]
\end{equation} where the symbol \(:\) denotes the full tensor contraction. This form of the solution is already simplified according to the assumptions of an isothermal and incompressible flow. For the full solution, one can refer to Chapter 4 in~\cite{struchtrup_macroscopic_2005}. The initial density profile is related to the initial pressure profile via the gas state equation \(p(0)=\rho(0) R T_0\). The equilibrium term, \(f_\alpha^{eq}\), can be calculated using Eq.~(\ref{feq}). We also note that the ``first-order" means the order in terms of the Knudsen number for the asymptotic Chapman-Enskog expansion. It has no bearing on the numerical accuracy in space and time.

For this particular problem, three turbulence quantities are often of interest, i.e., the integral of kinetic energy,

\begin{equation}
    E=\frac{1}{|\Omega|}\int_{\Omega} \frac{1}{2} \left(u^2+v^2+w^2\right)  \,d\bm{r},
\end{equation}
the kinetic energy dissipation rate \(\epsilon=-dE/dt\), and the integral of enstrophy,

\begin{equation}
\mathcal{Z}=\frac{1}{2|\Omega|}\int_{\Omega} \left[\left(v_x-u_y\right)^2+\left(u_z-w_x\right)^2+\left(w_y-v_z\right)^2\right]d\bm{r},
\end{equation} where \(\Omega\) denotes the periodic box domain.
For the present TGV problem, the kinetic energy dissipation rate can also be calculated as

\begin{equation}
    \epsilon=\frac{\mu_0 }{\rho_0 |\Omega|}\int_{\Omega} \left[ \left(u_y+v_x\right)^2+\frac{1}{2} \left(u_z+w_x\right)^2+\frac{1}{2} \left(v_z+w_y\right)^2+u_x^2+v_y^2+w_z^2\right]d\bm{r} \label{ratefromspace}
\end{equation} by using spatial derivatives.

To facilitate the comparison, we use the solution of the NS equations numerically solved by using a de-aliased pseudo-spectral code developed at University\'e catholique de Louvain with a low-storage three-step Runge-Kutta scheme for time integration. The data were generated using a mesh resolution of \(512^3\), and can be downloaded from \href{http://www.as.dlr.de/hiocfd/spectral_Re1600_512.gdiag}{the website of the 1st International Workshop on High-Order CFD Methods}~\cite{TGVBenchmarkData}. In the following figures, the data will be labeled as ``Ref''.

\section{Numerical analysis}\label{sec:numerical-results}
\subsection{Implementation}
The lattice Boltzmann scheme Eq.~(\ref{eq:scheme}) is implemented in a code based on the OPS library~\cite{Reguly_et_al_2018}, which is designed following the domain-specific language approach~\cite{HiLeMMSDSL}. The calculation of turbulence quantities is implemented as  ``user kernel" functions for the TGV application.  The code can be found at~\href{https://github.com/inmeso/mplb/tree/feature/TaylorGreanVortex}{Github}~\cite{mpblcode}.

During simulations, it is convenient to work with a non-dimensional system with \(\sqrt{R T_0}\) as the characteristic speed, and \(L\) as the reference length~\cite{meng_lattice_2013}. With such a non-dimensional system,  the mesh size, \(h\), and the time step, \(dt\), for each mesh resolution can be listed in Table~\ref{Tab:NumericalParameters}. It is worth noting that the Mach number is kept constant when the mesh is refined.

\begin{table}[h!]
    \begin{center}
        \begin{tabular}[]{|c|ccc|}
            \toprule
             & \(128^3\) & \(256^3 \) & \(512^3\)  \\
            \midrule
           \(h=2\pi/N\)   & 0.0494739  &  0.0246399          & 0.0122959  \\
            \(dt=h/\sqrt{3}\)   & 0.0285638  &0.0142259           & 0.00709902  \\
           \bottomrule
        \end{tabular}
        \caption{Mesh size and time step}
        \label{Tab:NumericalParameters}
    \end{center}
\end{table}
\subsection{Reconstruction schemes}\label{sec:methods-of-reconstruction}

To reconstruct the turbulence quantities from the velocity data, we use various orders of central difference schemes to evaluate the first-order derivative of a function, \(\varphi\), at a grid point \((i,j,k)\). Using \(\varphi_x\) as example, they are the second-order,

\[
    \frac{\varphi_{i+1}-\varphi_{i-1}}{2h},
\]
the fourth-order

\[
    \frac{-\varphi_{i+2}+8\varphi_{i+1}-8\varphi_{i-1}+\varphi_{i-2}}{12h},
\]
the sixth-order

\[
    \frac{\varphi_{i+3}-9\varphi_{i+2}+45\varphi_{i+1}-45\varphi_{i-1}+9\varphi_{i-2}-\varphi_{i-3}}{60h},
\]
and the eighth-order

\[
    \frac{1}{h}\left[\frac{1}{280}(\varphi_{i-4}-\varphi_{i+4})+\frac{4}{105}(\varphi_{i+3}-\varphi_{i-3})+\frac{1}{5}(\varphi_{i-2}+\varphi_{i+2})+\frac{4}{5}(\varphi_{i+1}-\varphi_{i-1})\right].
\] In these formulations, we omit the index \(j\) and \(k\) for convenience, which are not changed for calculating the derivative in the \(x\) direction. Moreover, we do not need to specifically treat the boundary points for this periodic problem.

During simulations, we do not store velocity data if they are not required for further analysis. Instead, we implement the reconstruction process into the code by ``user kernel" functions, and just store the turbulence quantities.

The kinetic energy dissipation rate can be also be reconstructed from the time series of kinetic energy. For this purpose, we use a third-order spline interpolation. This is implemented by using the ``interp1d" function provided by the open-source SciPy package. Then, the time derivative is evaluated by a central difference scheme, which is also implemented by using the ``derivative" function of SciPy. This procedure is particularly useful as we cannot align the LBM timesteps with those of the pseudo-spectral data. Due to the tie to the space step, the LBM timestep can only be those as shown in Table ~\ref{Tab:NumericalParameters} but the timestep of pseudo-spectral data with a resolution of \(512^3\) is \(0.01\).


To assess the accuracy, the \(L^2\) norm is utilized. Considering a vector \(S\) for results and a vector \(\mathcal{C}\) for ``correct'' values, the error \(\sigma\) is calculated as

\begin{equation}
    \sigma=\sqrt{\frac{(S-\mathcal{C})\cdot(S-\mathcal{C})}{\mathcal{C}\cdot \mathcal{C}}}.
\end{equation}
It will be applied to calculating the error of kinetic energy \(\sigma_{E}\), the error of kinetic energy dissipation rate~\(\sigma_{\epsilon}\), and the error of enstrophy \(\sigma_\mathcal{Z}\). The ``correct'' solution vectors are those of the pseudo-spectral data.

\subsection{Kinetic energy}\label{sec:kinetic-energy}

\begin{figure}
    \begin{center}
        \includegraphics[width=0.65\textwidth]{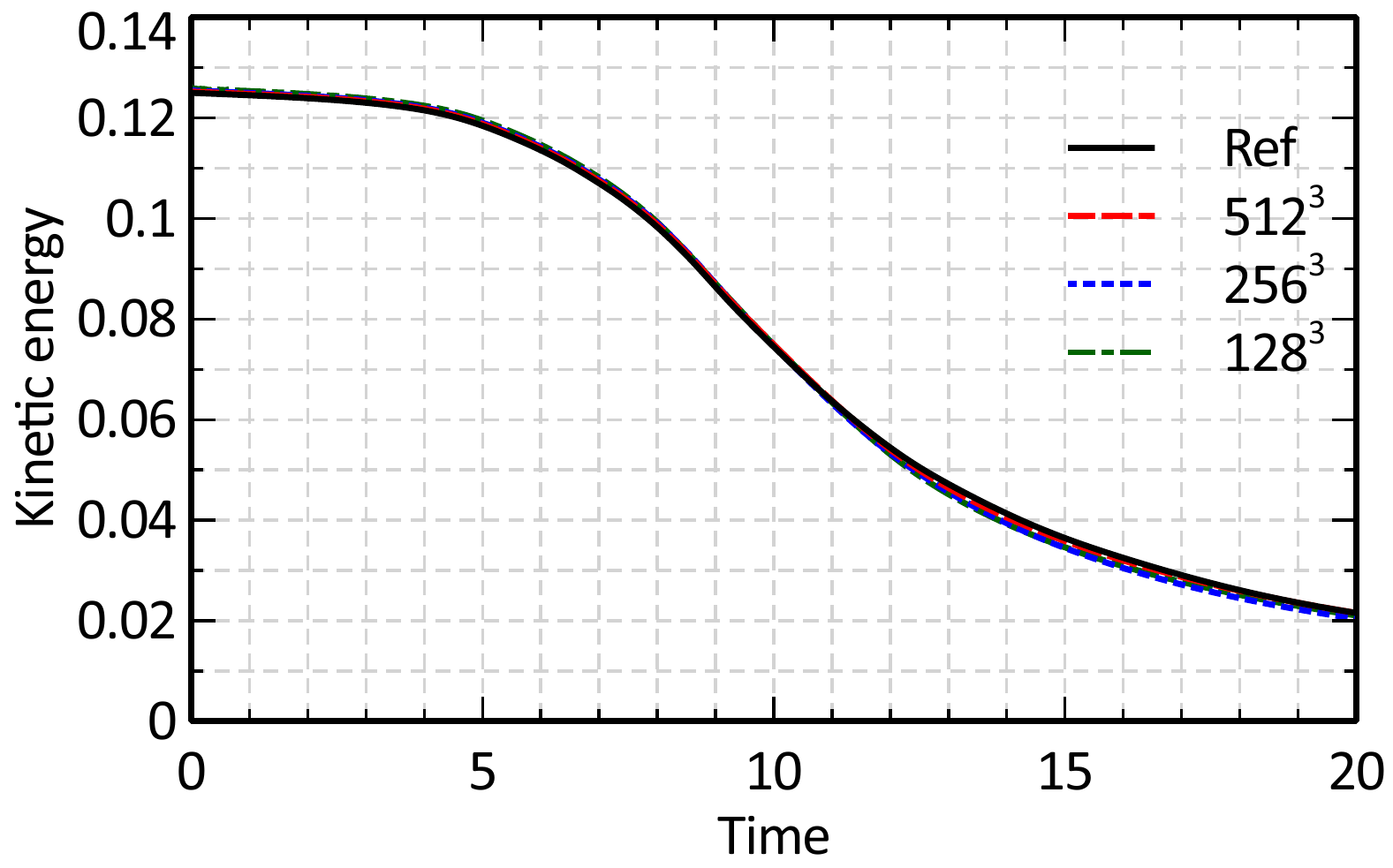}
        \caption{Predicted profiles of kinetic energy. The errors (\(\sigma_E\)) are \(1.3729\%\), \(1.3735\%\) and \(0.49827\%\) with the mesh resolutions of \(128^3\), \(256^3\) and \(512^3\), respectively. \label{fig:kinetictenergy}}
    \end{center}
\end{figure}

The simulations predict kinetic energy with considerably high accuracy, see Fig.~\ref{fig:kinetictenergy}. The error is already less than \(2\%\) with a mesh resolution of \(128^3\). Interestingly, a higher resolution of \(256^3\) does not improve the accuracy. A similar observation was also found in \cite{Mimeau2021}, cf. Fig.~(11). However, it will be shown later that the resolution of \(256^3\) does improve the prediction of the kinetic energy dissipation rate.

\subsection{Kinetic energy dissipation}\label{sec:kinetic-energy-dissipation}
\begin{figure}
    \begin{center}
        \includegraphics[width=0.65\textwidth]{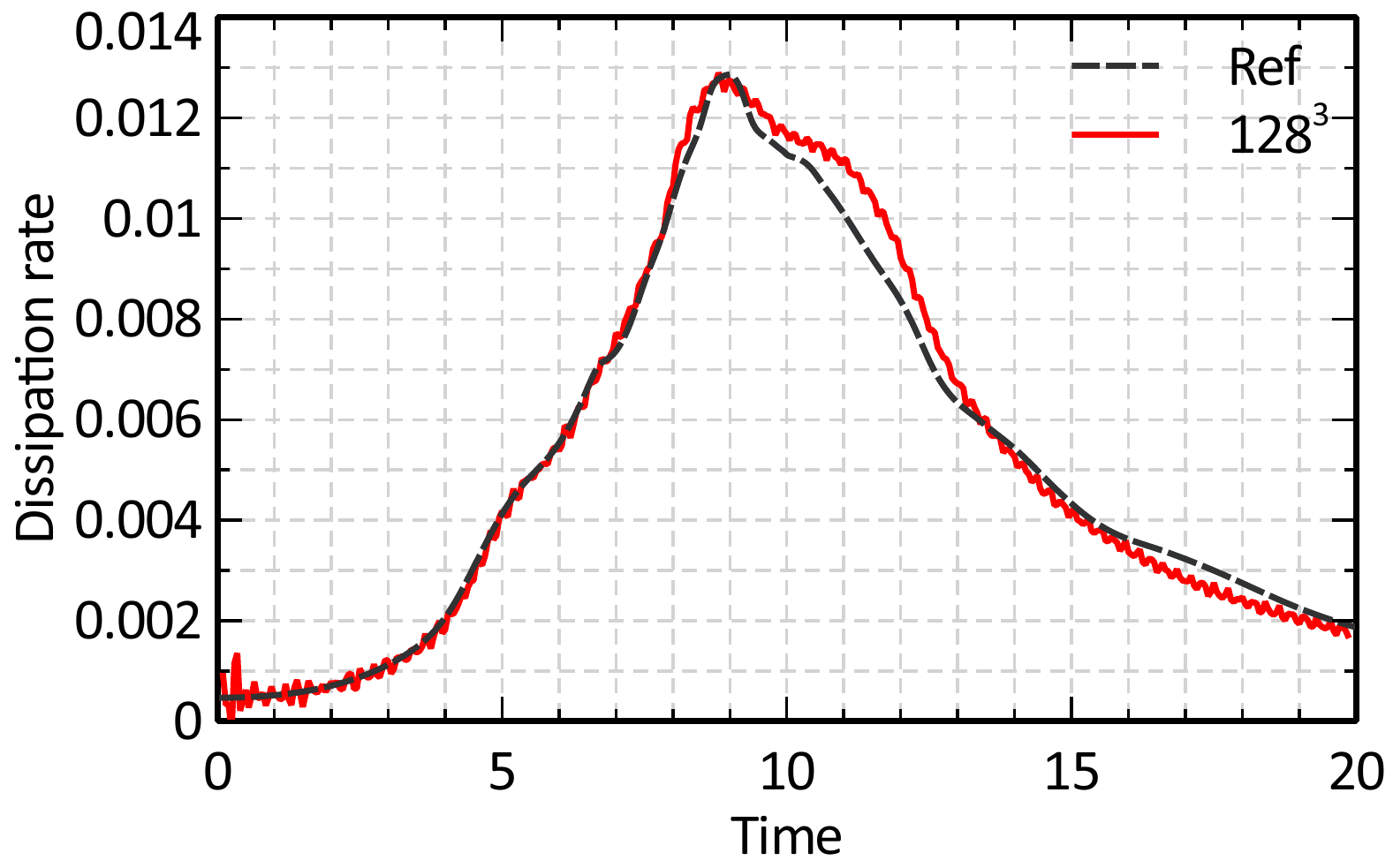}
        \includegraphics[width=0.65\textwidth]{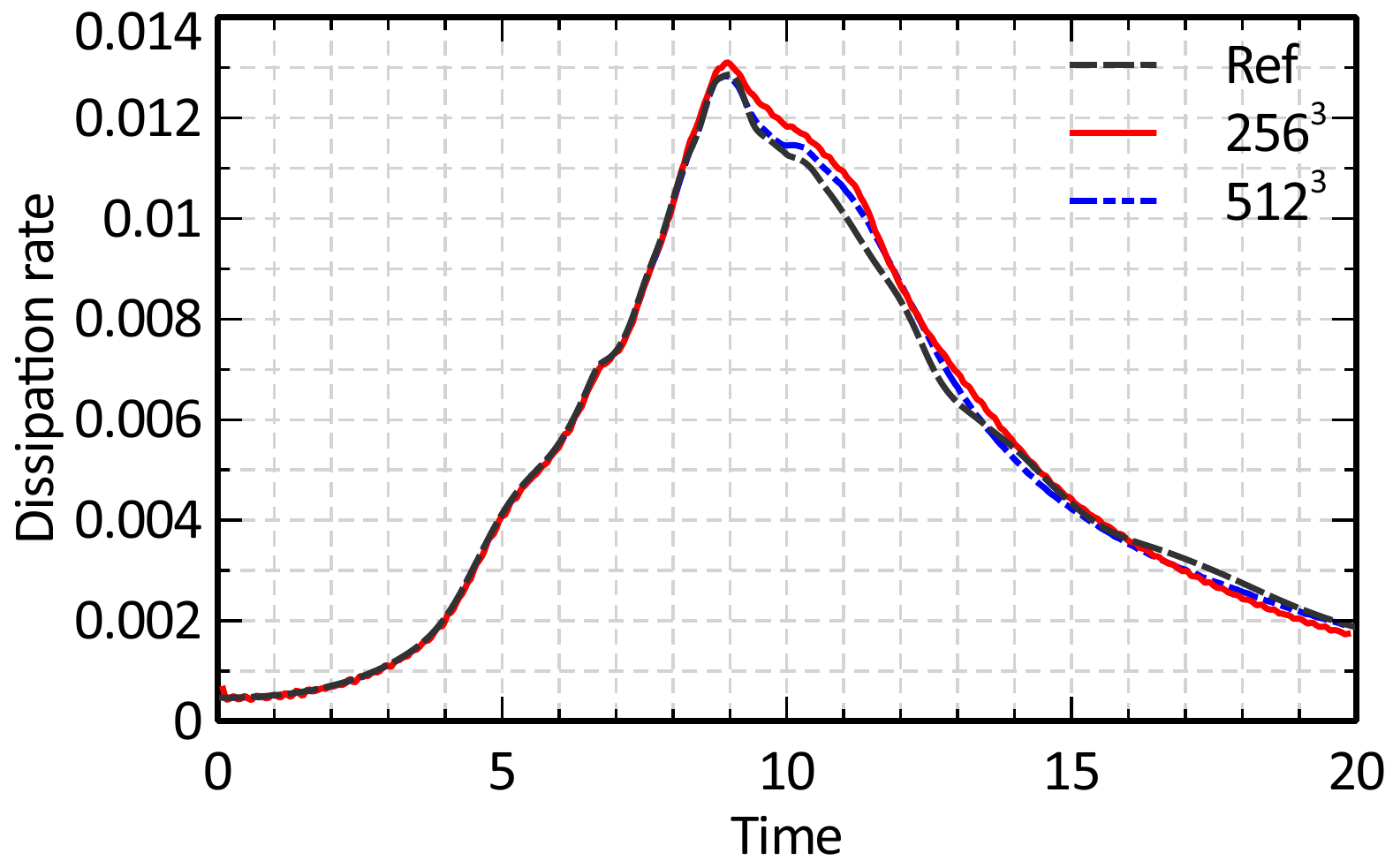}
        \caption{Kinetic energy dissipation rate reconstructed from time series.}
        \label{fig:dissipationFromTime}
    \end{center}
\end{figure}

We first focus on the reconstruction of the kinetic energy dissipation rate from the time series of kinetic energy. For this purpose, we use the second-order central difference scheme for \(-dE/dt\) from the kinetic energy data between \(t=0.1\) and \(t=19.9\) with a step \(dt=0.05\), which are passed to the ``derivative" function of SciPy. The obtained data are plotted in Fig.~\ref{fig:dissipationFromTime}. It can seen that there are significant wiggles from the results of \(128^3\). This is also observed in~\cite{Mimeau2021} for both LBM predictions and those produced by the semi-Lagrangian vortex method for solving the NS equations using a low resolution mesh.

The errors are also listed in Table~\ref{Tab:dissipationErrorTime}. The results tend to agree with the pseudo-spectral data well. Moreover, it is found that increasing the order of the central difference scheme for the  discretization of \(-dE/dt\)  does not help to improve the accuracy.

\begin{figure}
    \begin{center}
        \includegraphics[width=0.65\textwidth]{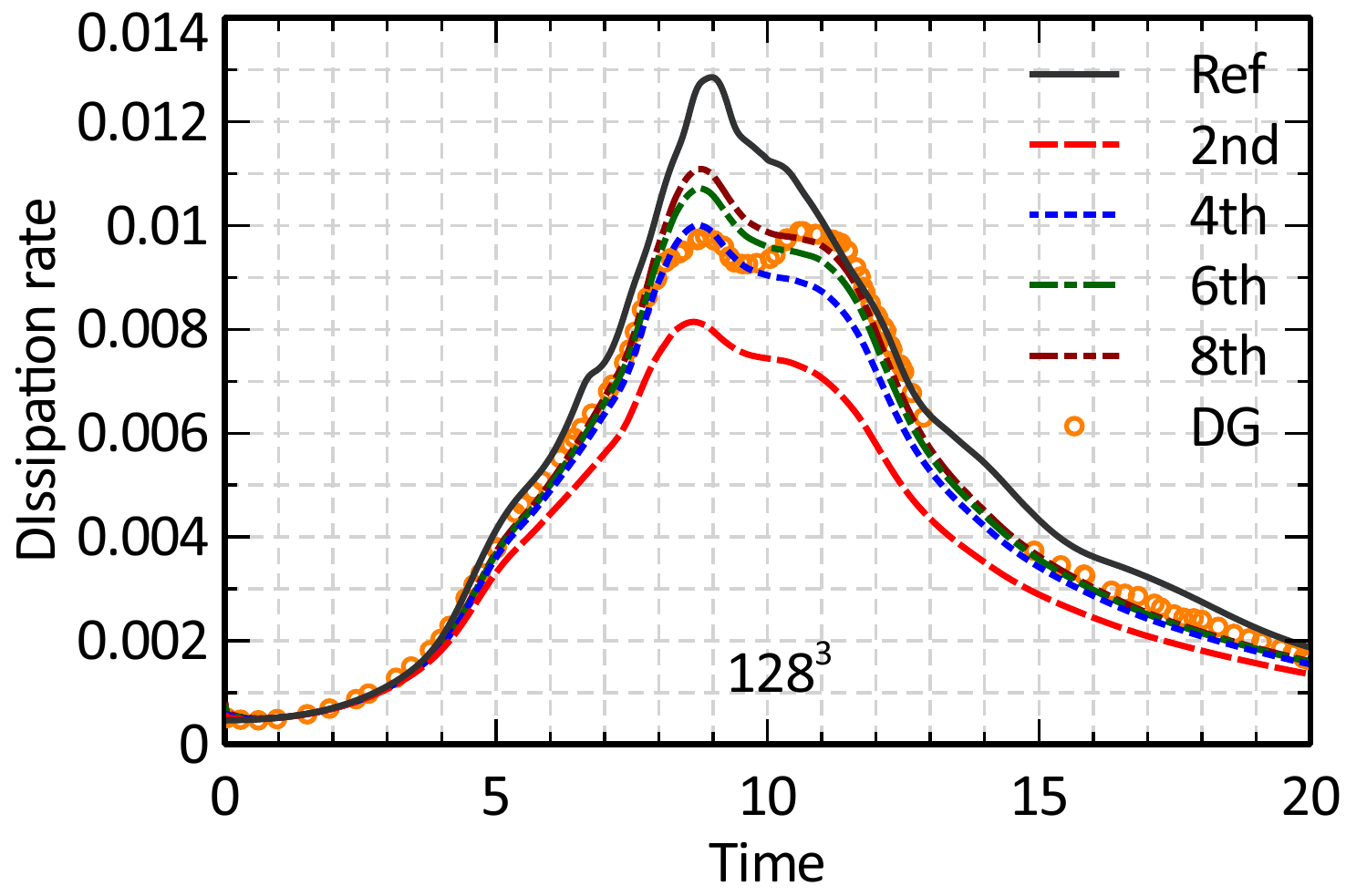}
        \includegraphics[width=0.65\textwidth]{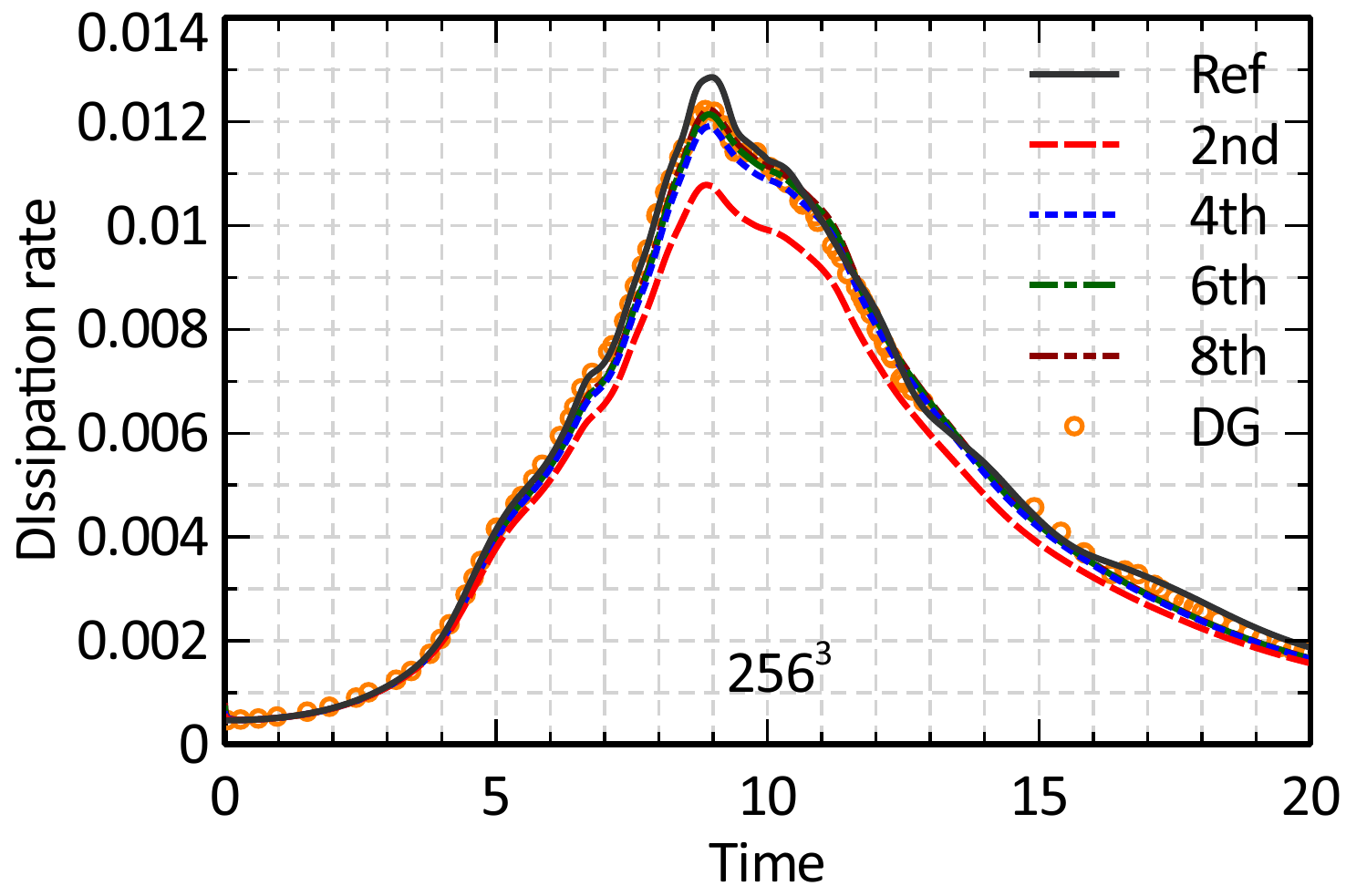}
        \includegraphics[width=0.65\textwidth]{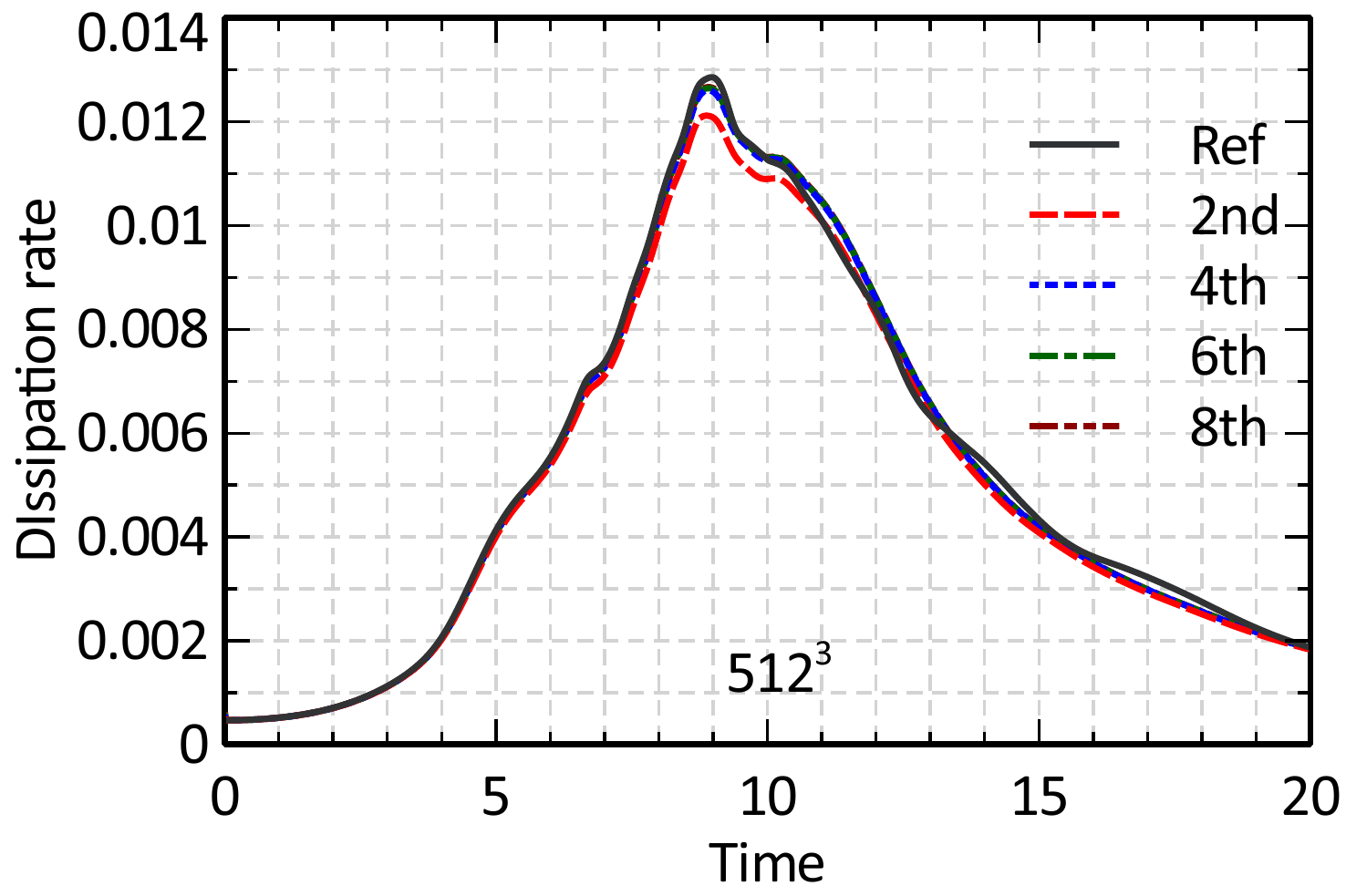}
        \caption{Kinetic energy dissipation rate reconstructed from spatial derivatives.}
        \label{fig:dissipationFromSpace}
    \end{center}
\end{figure}

\begin{table}[h!]
    \begin{center}
        \begin{tabular}[]{|c|ccc|}
            \toprule
                                             & \(128^3\)  & \(256^3 \) & \(512^3\)  \\
            \midrule
            Error(\(\sigma_{\epsilon}\)) & \(6.36\%\) & \(4.70\%\) & \(2.97\%\) \\
            \bottomrule
        \end{tabular}
        \caption{Error of kinetic energy dissipation rate reconstructed from time series}
        \label{Tab:dissipationErrorTime}
    \end{center}
\end{table}

Next we consider the kinetic energy dissipation rate reconstructed using spatial derivatives, i.e. Eq.~(\ref{ratefromspace}). For this purpose, we use four central difference schemes to reconstruct \(\epsilon\). The obtained profiles are plotted in Fig.~\ref{fig:dissipationFromSpace} while the errors are presented in Table~\ref{Tab:dissipationErrorSpace}.

It is shown that the high-order central difference schemes can significantly improve the accuracy when the mesh resolution is low. For both \(128^3\) and \(256^3\), the errors of the eighth-order scheme is nearly \(3\) times lower than those of the second-order scheme. For \(512^3\), the second-order scheme can also achieve an error less than \(5\%\) while the predictions of other three higher-order schemes show trend of convergence. Compared with the reconstruction from the time series, the results of \(512^3\) and \(256^3\) start to match the accuracy from the fourth-order and sixth-order schemes, respectively. For \(128^3\), the prediction by using a spatial derivative appears unable to match the accuracy with the tested central difference scheme. The present finding also explains the inconsistent observation in the accuracy of predicting the kinetic energy dissipation rate and enstrophy in~\cite{Mimeau2021}.

To further evaluate the accuracy of LBM, we also include the results of a fourth-order discontinuous-Galerkin (DG) finite-element scheme for the compressible NS equations in Fig.~\ref{fig:dissipationFromSpace}. The results of DG are obtained by digitizing Fig.1 in \cite{dg4th}. Due to the legend of Fig.1 in \cite{dg4th}, the data between $t=13$ and $t=15$ are missing but this should not have great impact.  From the comparison, we can see that, with the higher-order reconstruction scheme, the LBM predicts the dissipation rate with accuracy comparable to the fourth-order DG scheme for the NS equations.

We also calculate the order of accuracy (i.e., the rate of convergence to the ``correct" solution) demonstrated by Eq.~(\ref{eq:scheme}) during the numerical experiments with different mesh resolution as

\begin{equation}
 \mathcal{O}=\frac{\log\frac{\sigma_{N_c}}{\sigma_{N_r}}}{\log\frac{N_r}{N_c}}
\end{equation} based on the calculated error \(E\) with a finer mesh resolution \(N_r\)
and a coarser mesh resolution \(N_c\). The values are listed in Table~\ref{Tab:slopeMesh}. With a second-order reconstruction, the lattice Boltzmann simulation demonstrates an average order of accuracy \(1.46\), while the order is reduced to \(1.05\) if using an eighth-order reconstruction. This is perhaps due to the fact that the results with the eighth-order reconstruction are already fairly accurate with a mesh resolution of \(256^3\).

Overall, LBM simulations tend to present very accurate kinetic energy dissipation rate. This should be due to its features of low numerical dissipation error and exactly satisfying mass and momentum conservation. The observations suggest the necessity of using schemes at least of fourth-order accuracy for evaluating spatial derivatives in processing LBM simulation results apart from its formal second-order accuracy relative to Eq.~(\ref{pde}).

\begin{table}[h!]
    \begin{center}
        \begin{tabular}[]{|c|ccc|}
            \toprule
            Error(\(\sigma_{\epsilon}\)) & \(128^3\)   & \(256^3 \)  & \(512^3\)   \\
            \midrule
            2nd                              & \(31.54\%\) & \(12.29\%\) & \(4.184\%\) \\
            4th                              & \(17.90\%\) & \(5.236\%\) & \(2.653\%\) \\
            6th                              & \(13.43\%\) & \(4.226\%\) & \(2.649\%\) \\
            8th                              & \(11.33\%\) & \(3.937\%\) & \(2.644\%\) \\
            \bottomrule
        \end{tabular}
        \caption{Errors of kinetic energy dissipation rate reconstructed from spatial derivatives}
        \label{Tab:dissipationErrorSpace}
    \end{center}
\end{table}

\begin{table}[h!]
    \begin{center}
        \begin{tabular}[]{|c|ccc|}
            \toprule
            Order  & \(256^3 \) & \(512^3\) & Average \\
            \midrule
            2nd       & 1.36       & 1.55      & 1.46    \\
            4th         & 1.77       & 0.98      & 1.38    \\
            6th     & 1.67       & 0.67      & 1.17    \\
            8th          & 1.53       & 0.57      & 1.05    \\
            \bottomrule
        \end{tabular}
        \caption{Order of accuracy with various mesh resolutions and reconstruction schemes}
        \label{Tab:slopeMesh}
    \end{center}
\end{table}

\subsection{Enstrophy}\label{sec:enstrophy}
\begin{figure}
    \begin{center}
        \includegraphics[width=0.65\textwidth]{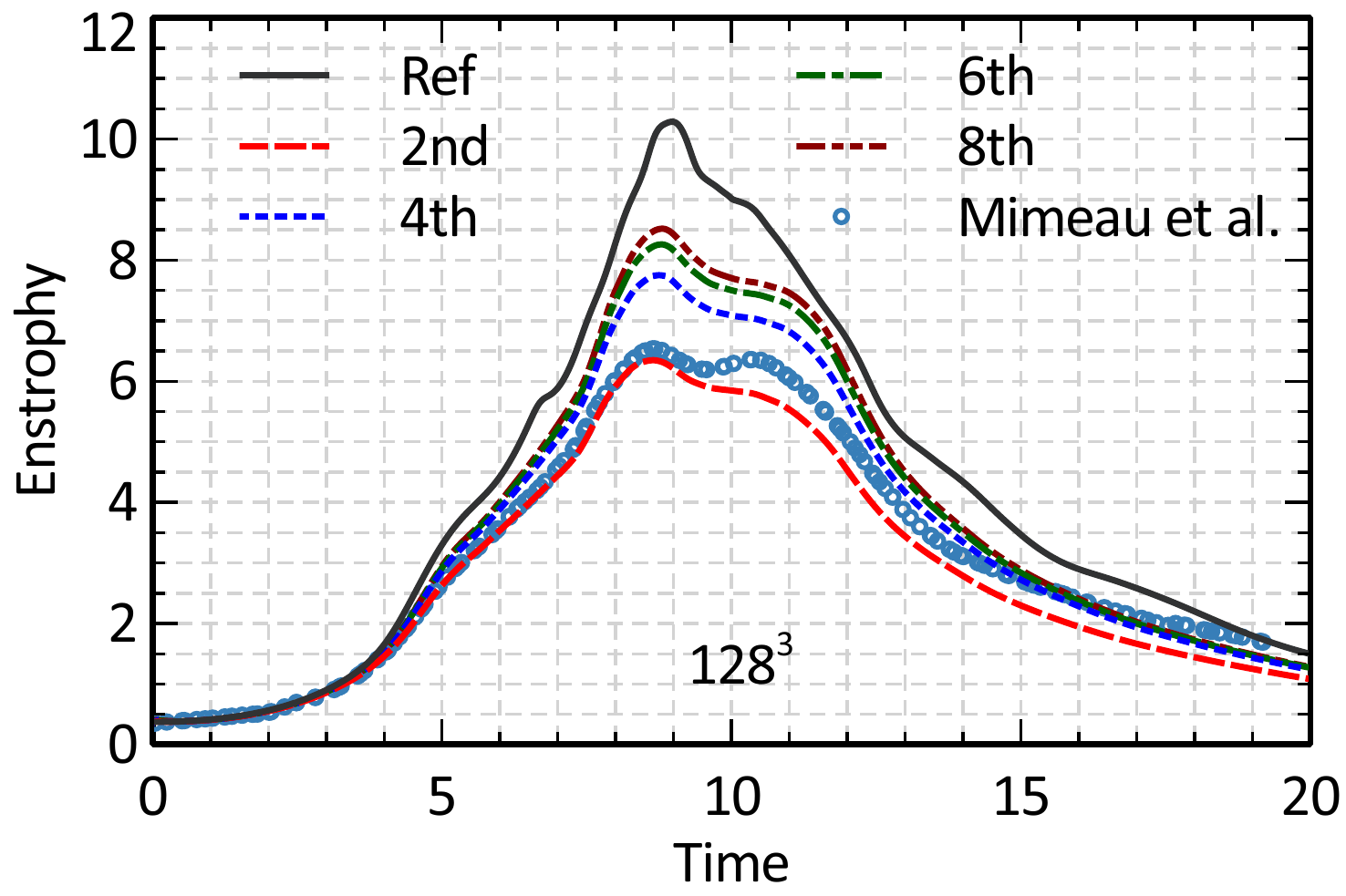}
        \includegraphics[width=0.65\textwidth]{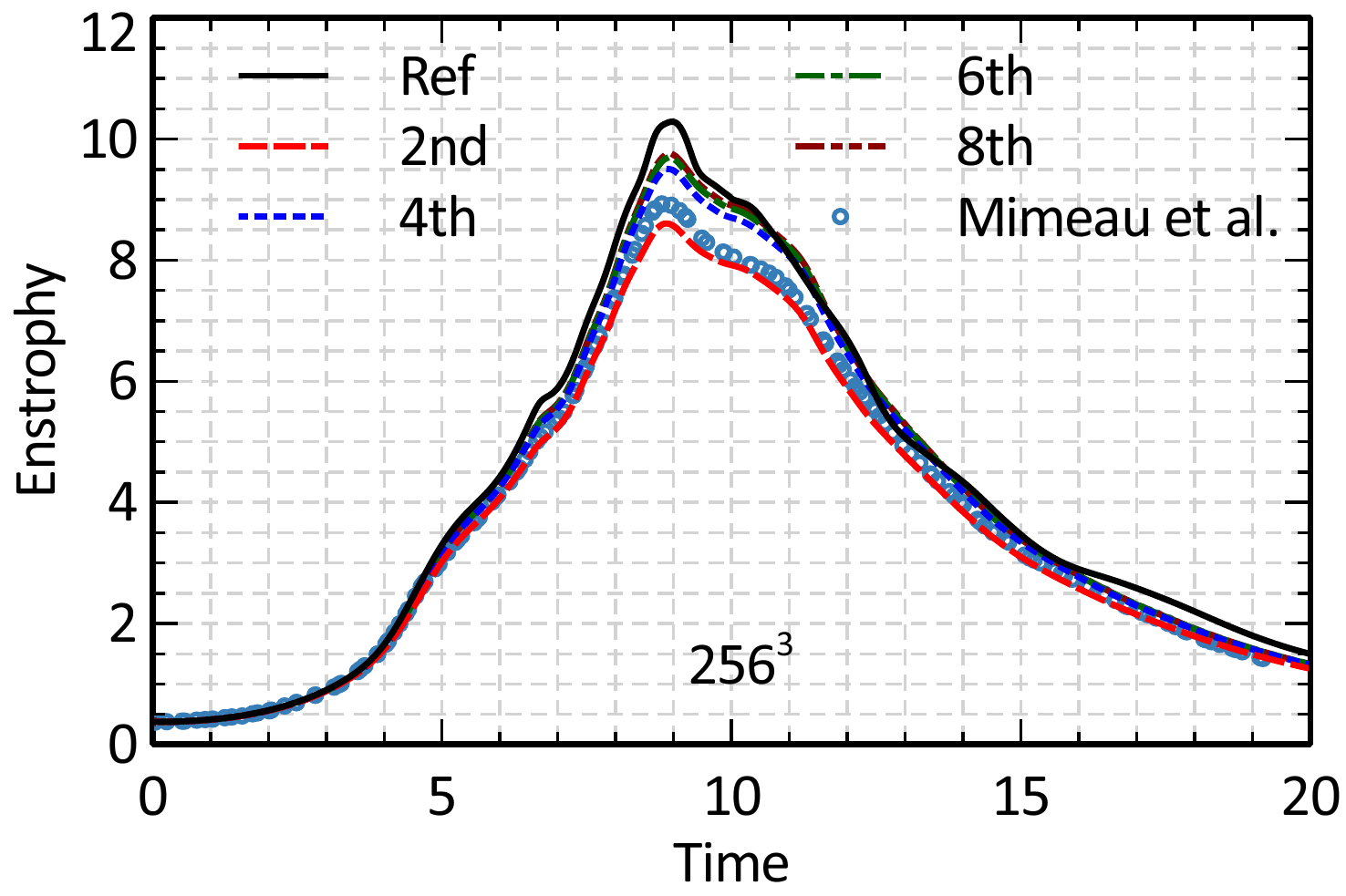}
        \includegraphics[width=0.65\textwidth]{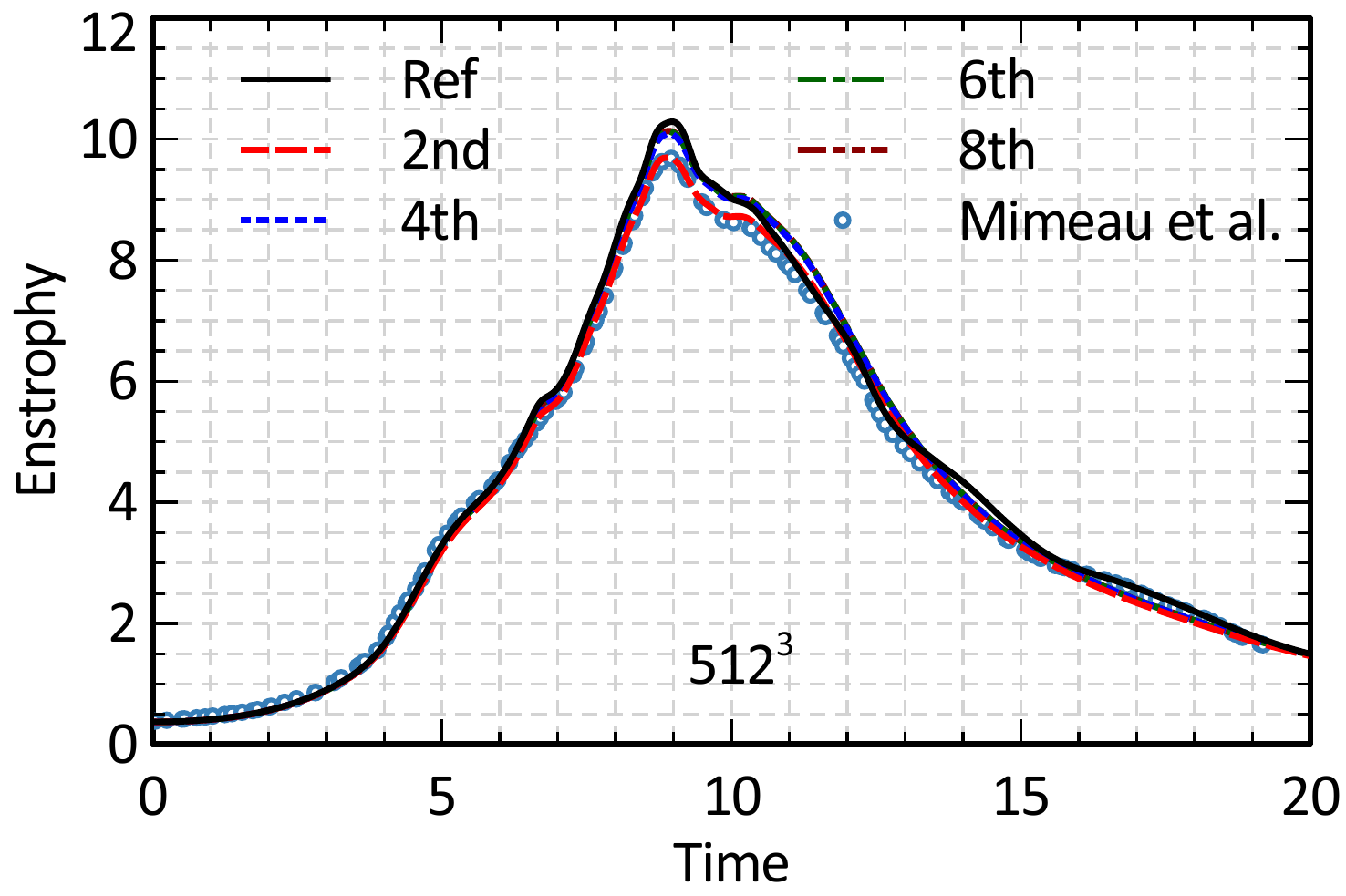}
        \caption{Kinetic energy dissipation rate reconstructed from spatial derivatives.}
        \label{fig:enstrophy}
    \end{center}
\end{figure}

The enstrophy can only be reconstructed from spatial derivatives, which is proportional to the kinetic energy dissipation rate for this problem. Therefore, we observe similar accuracy for this quantity, see Fig.~\ref{fig:enstrophy} and Table~\ref{tab:enstrophy}.

\begin{table}[h!]
    \begin{center}
        \begin{tabular}[]{|c|ccc|}
    \toprule
    Error(\(\sigma_{\mathcal{Z}}\)) &       \(128^3\) &       \(256^3\) &       \(512^3\) \\
    \midrule
    2nd &  32.70\% &  12.42\% &  4.19\% \\
    4th &  19.46\% &  5.32\% &  2.65 \%\\
    6th &  15.25\% &  4.30 \%&  2.65 \%\\
    8th &  13.29\% &  4.01\% &  2.65 \%\\
    \bottomrule
    \end{tabular}
    \caption{Errors of enstrophy reconstructed from spatial derivatives}
    \label{tab:enstrophy}
\end{center}
\end{table}
In Fig.~\ref{fig:enstrophy}, there are comparisons with the results obtained from the multiple relaxation time model by Mimeau \textit{et al}. \cite{Mimeau2021}.  Apart from the larger difference for the case of \(128^3\), perhaps due to the different collision terms, the present results reconstructed using the second-order scheme agree in general with those of Mimeau \textit{et al}., which were also reconstructed using a second-order scheme. In particular, they agree with each other well at \(512^3\). Thus, the present simulations also confirm previous validations on this problem. Moreover, this further acknowledges the importance of using higher-order reconstruction schemes which significantly improve the results in comparison with those of the second-order reconstruction.

\section{Concluding remarks}\label{sec:concluding-remarks}

We have studied the impact of reconstruction schemes on interpreting LBM simulation results. The classical Taylor-Green vortex problem is chosen for the purpose. Simulations are conducted for the well-documented case at  Reynolds number of \(1600\). It is found that using high-order central difference schemes can significantly improve the prediction of the kinetic energy dissipation rate and enstrophy using spatial derivatives. This is particularly true with a low resolution mesh \(128^3\) where the eighth-order scheme can still reduce \(\sim 2\%\) of error in comparison to the sixth-order scheme.   The most significant improvement happens between the fourth-order and the second-order schemes where the error is reduced by about half for all three meshes. However, there is no significant improvement among fourth-order, sixth-order and eighth-order schemes for \(256^3\) and \(512^3\). We also studied the reconstruction of the kinetic energy dissipation rate from the kinetic energy time series. It is shown that a second-order scheme can already reconstruct the kinetic energy dissipation rate well.

Overall, the LBM simulations demonstrates excellent accuracy for this particular problem considering its second-order accuracy relative to the Boltzmann transport equation.  In fact, it demonstrates similar accuracy to  a fourth-order discontinuous-Galerkin finite-element scheme for the Navier-Stokes equations. This should be attributed to its features of very low numerical diffusion and satisfying mass and momentum conservation at the mesoscopic level. This observation is very encouraging since its algorithm is very simple and requires only the computational cost of a typical first-order numerical scheme.  Moreover, the Poisson equation is not needed for simulating incompressible flows.

Our observation suggests that high-order reconstruction is necessary for extracting non-primitive flow quantities concerning spatial derivatives from LBM simulations. Based on Tables \ref{Tab:dissipationErrorSpace} and \ref{tab:enstrophy}, a fourth-order reconstruction can balance the accuracy and the computational expense if such quantities are needed dynamically. For example, evaluating the fluid force on moving particles immersed in the fluid. On the other hand, the good accuracy of reconstruction from time series favors extracting the fluid force through the momentum exchange method~\cite{Ladd1994}. It is not only a convenient numerical technique but could also have higher accuracy over calculating spatial derivatives.

\section*{Acknowledgments}

This work was conducted under the support of the Computational Science Centre for Research Communities, and the EPSRC grants EP/P022243/1, EP/T026170/1 and EP/X035875/1. The simulations were performed by utilizing both nodes of the STFC Cloud Service and the ARCHER2 machine.

We also thank Dr.~Jian Fang for providing and pointing out the source of the pseudo-spectral data.

\bibliographystyle{unsrtnat}

\begin{thebibliography}{22}
	\providecommand{\natexlab}[1]{#1}
	\providecommand{\url}[1]{\texttt{#1}}
	\expandafter\ifx\csname urlstyle\endcsname\relax
	  \providecommand{\doi}[1]{doi: #1}\else
	  \providecommand{\doi}{doi: \begingroup \urlstyle{rm}\Url}\fi

	\bibitem[Ekaterinaris(2005)]{Ekaterinaris2005}
	John~A. Ekaterinaris.
	\newblock High-order accurate, low numerical diffusion methods for
	  aerodynamics.
	\newblock \emph{Prog. Aerosp. Sci.}, 41\penalty0 (3):\penalty0 192--300, 2005.
	\newblock ISSN 0376-0421.
	\newblock \doi{https://doi.org/10.1016/j.paerosci.2005.03.003}.
	\newblock URL
	  \url{https://www.sciencedirect.com/science/article/pii/S0376042105000473}.

	\bibitem[Chen and Doolen(1998)]{chen_lattice_1998}
	Shiyi Chen and Gary~D Doolen.
	\newblock Lattice {Boltzmann} method for fluid flows.
	\newblock \emph{Annu. Rev. Fluid Mech.}, 30\penalty0 (1):\penalty0 329--364,
	  1998.

	\bibitem[Wolfram(1986)]{Wolfram1986}
	Stephen Wolfram.
	\newblock Cellular automaton fluids 1: Basic theory.
	\newblock \emph{J. Statist. Phys.}, 45\penalty0 (3):\penalty0 471--526, 1986.
	\newblock ISSN 1572-9613.
	\newblock \doi{10.1007/BF01021083}.
	\newblock URL \url{https://doi.org/10.1007/BF01021083}.

	\bibitem[Frisch et~al.(1986)Frisch, Hasslacher, and
	  Pomeau]{frisch_lattice-gas_1986}
	U.~Frisch, B.~Hasslacher, and Y.~Pomeau.
	\newblock Lattice-{Gas} {Automata} for the {Navier}-{Stokes} {Equation}.
	\newblock \emph{Phys. Rev. Lett.}, 56\penalty0 (14):\penalty0 1505--1508, April
	  1986.
	\newblock \doi{10.1103/PhysRevLett.56.1505}.

	\bibitem[He and Luo(1997{\natexlab{a}})]{he_priori_1997}
	Xiaoyi He and L.S. Luo.
	\newblock A priori derivation of the lattice {Boltzmann} equation.
	\newblock \emph{Phys. Rev. E}, 55\penalty0 (6):\penalty0 6333--6336,
	  1997{\natexlab{a}}.
	\newblock \doi{10.1103/PhysRevE.55.R6333}.

	\bibitem[He and Luo(1997{\natexlab{b}})]{he_theory_1997}
	Xiaoyi He and Li-Shi Luo.
	\newblock Theory of the lattice {Boltzmann} method: {From} the {Boltzmann}
	  equation to the lattice {Boltzmann} equation.
	\newblock \emph{Phys. Rev. E}, 56:\penalty0 6811--6817, 1997{\natexlab{b}}.

	\bibitem[Shan and He(1998)]{shan_discretization_1998}
	Xiaowen Shan and Xiaoyi He.
	\newblock Discretization of the velocity space in the solution of the
	  {Boltzmann} equation.
	\newblock \emph{Phys. Rev. Lett.}, 80\penalty0 (1):\penalty0 65--68, January
	  1998.

	\bibitem[Shan et~al.(2006)Shan, Yuan, and Chen]{shan_kinetic_2006}
	X~W Shan, X~F Yuan, and H~D Chen.
	\newblock Kinetic theory representation of hydrodynamics: {A} way beyond the
	  {N}avier {S}tokes equation.
	\newblock \emph{J. Fluid Mech.}, 550:\penalty0 413--441, 2006.

	\bibitem[Aidun et~al.(2010)Aidun, Clausen, and
	  Woodruff]{aidun_lattice-boltzmann_2010}
	C.~K. Aidun, J.~R. Clausen, and G.~W. Woodruff.
	\newblock Lattice-{B}oltzmann method for complex flows.
	\newblock \emph{Annu. Rev. Fluid Mech.}, 42\penalty0 (1):\penalty0 439--72,
	  2010.
	\newblock ISSN 0066-4189{\textbackslash}r1545-4479.
	\newblock \doi{10.1146/annurev-fluid-121108-145519}.

	\bibitem[Marié et~al.(2009)Marié, Ricot, and Sagaut]{marie_comparison_2009}
	Simon Marié, Denis Ricot, and Pierre Sagaut.
	\newblock Comparison between lattice {Boltzmann} method and {Navier}–{Stokes}
	  high order schemes for computational aeroacoustics.
	\newblock \emph{J. Comput. Phys.}, 228\penalty0 (4):\penalty0 1056--1070, March
	  2009.
	\newblock ISSN 0021-9991.
	\newblock \doi{10.1016/j.jcp.2008.10.021}.
	\newblock URL
	  \url{https://www.sciencedirect.com/science/article/pii/S002199910800538X}.

	\bibitem[Mimeau et~al.(2021)Mimeau, Marié, and Mortazavi]{Mimeau2021}
	Chloé Mimeau, Simon Marié, and Iraj Mortazavi.
	\newblock A comparison of semi-{L}agrangian vortex method and lattice
	  {B}oltzmann method for incompressible flows.
	\newblock \emph{Comput. \& Fluids}, 224:\penalty0 104946, June 2021.
	\newblock ISSN 0045-7930.
	\newblock \doi{10.1016/j.compfluid.2021.104946}.
	\newblock URL
	  \url{https://www.sciencedirect.com/science/article/pii/S0045793021001134}.

	\bibitem[Haussmann et~al.(2019)Haussmann, Simonis, Nirschl, and
	  Krause]{Haussmann2019}
	Marc Haussmann, Stephan Simonis, Hermann Nirschl, and Mathias~J. Krause.
	\newblock Direct numerical simulation of decaying homogeneous isotropic
	  turbulence — numerical experiments on stability, consistency and accuracy
	  of distinct lattice boltzmann methods.
	\newblock \emph{Int. J. Modern Phys. C}, 30\penalty0 (09):\penalty0 1950074,
	  2019.
	\newblock \doi{10.1142/S0129183119500748}.
	\newblock URL \url{https://doi.org/10.1142/S0129183119500748}.

	\bibitem[Nathen et~al.(2018)Nathen, Gaudlitz, Krause, and Adams]{Nathen2018}
	Patrick Nathen, Daniel Gaudlitz, Mathias~J. Krause, and Nikolaus~A. Adams.
	\newblock On the stability and accuracy of the {BGK}, {MRT} and {RLB}
	  {B}oltzmann schemes for the simulation of turbulent flows.
	\newblock \emph{Commun. Comput. Phys.}, 23\penalty0 (3), 2018.
	\newblock ISSN 18152406.
	\newblock \doi{10.4208/cicp.OA-2016-0229}.
	\newblock URL
	  \url{http://www.global-sci.com/intro/article_detail/cicp/10551.html}.

	\bibitem[He et~al.(1998)He, Chen, and Doolen]{he_novel_1998}
	Xiaoyi He, Shiyi Chen, and Gary~D Doolen.
	\newblock A novel thermal model for the lattice {Boltzmann} method in
	  incompressible limit.
	\newblock \emph{J. Comput. Phys.}, 146\penalty0 (1):\penalty0 282--300, 1998.

	\bibitem[Succi(2018)]{Succi2018}
	Sauro Succi.
	\newblock \emph{The Lattice Boltzmann Equation: For Complex States of Flowing
	  Matter}.
	\newblock Oxford University Press, Oxford, 2018.
	\newblock \doi{10.1093/oso/9780199592357.001.0001}.
	\newblock URL
	  \url{https://oxford.universitypressscholarship.com/10.1093/oso/9780199592357.001.0001/oso-9780199592357}.

	\bibitem[Struchtrup(2005)]{struchtrup_macroscopic_2005}
	Henning Struchtrup.
	\newblock \emph{Macroscopic {Transport} {Equations} for {Rarefied} {Gas}
	  {Flows}}.
	\newblock Interaction of {Mechanics} and {Mathematics}. Springer Berlin /
	  Heidelberg, 2005.

	\bibitem[TGV()]{TGVBenchmarkData}
	URL \url{http://www.as.dlr.de/hiocfd/spectral_Re1600_512.gdiag}.

	\bibitem[Reguly et~al.(2018)Reguly, Mudalige, and Giles]{Reguly_et_al_2018}
	István~Z. Reguly, Gihan~R. Mudalige, and Michael~B. Giles.
	\newblock Loop tiling in large-scale stencil codes at run-time with ops.
	\newblock \emph{IEEE Trans. Parallel Distrib. Syst.}, 29\penalty0 (4):\penalty0
	  873--886, 2018.
	\newblock \doi{10.1109/TPDS.2017.2778161}.

	\bibitem[HiL()]{HiLeMMSDSL}
	URL \url{https://github.com/inmeso/hilemms}.

	\bibitem[mpb()]{mpblcode}
	URL \url{https://github.com/inmeso/mplb/tree/feature/TaylorGreanVortex}.

	\bibitem[Meng et~al.(2013)Meng, Zhang, Hadjiconstantinou, Radtke, and
	  Shan]{meng_lattice_2013}
	Jianping Meng, Yonghao Zhang, Nicolas~G Hadjiconstantinou, Gregg~A Radtke, and
	  Xiaowen Shan.
	\newblock Lattice ellipsoidal statistical {BGK} model for thermal
	  non-equilibrium flows.
	\newblock \emph{J. Fluid Mech.}, 718:\penalty0 347--370, 2013.


\bibitem{dg4th}
L.~T.~Diosady, S.~M.~Murman, Case 3.3: Taylor-Green vortex evolution, NASA Ames Research Center, 2015. \url{ https://www1.grc.nasa.gov/wp-content/uploads/C3.3_Ames.pdf}

	\bibitem[Ladd(1994)]{Ladd1994}
	Anthony J.~C. Ladd.
	\newblock Numerical simulations of particulate suspensions via a discretized
	  boltzmann equation. part 1. theoretical foundation.
	\newblock \emph{J. Fluid Mech.}, 271:\penalty0 285--309, 1994.
	\newblock ISSN 0022-1120.
	\newblock \doi{10.1017/S0022112094001771}.

	\end{thebibliography}

\end{document}